# Photochemistry and Haze Formation

## K. E. Mandt, A. Luspay-Kuti, A. Cheng
*The Johns Hopkins University Applied Physics Laboratory*

## K.-L. Jessup
*Southwest Research Institute*

## P. Gao
*University of California Berkeley*

One of the many exciting revelations of the New Horizons flyby of Pluto was the observation of global haze layers at altitudes as high as 200 km in the visible wavelengths. This haze is produced in the upper atmosphere through photochemical processes, similar to the processes in Titan's atmosphere. As the haze particles grow in size and descend to the lower atmosphere, they coagulate and interact with the gases in the atmosphere through condensation and sticking processes that serve as temporary and permanent loss processes. New Horizons observations confirm studies of Titan haze analogs suggesting that photochemically produced haze particles harden as they grow in size. We outline in this chapter what is known about the photochemical processes that lead to haze production and outline feedback processes resulting from the presence of haze in the atmosphere, connect this to the evolution of Pluto's atmosphere, and discuss open questions that need to be addressed in future work.

## 1. INTRODUCTION

The New Horizons flyby of Pluto revealed global haze layers extending up to altitudes as high as 200 km in visible images as discussed in *Cheng et al.* (2017). Although early observations of Pluto's atmosphere from groundbased and Earth-orbiting telescopes indicated that haze may be present, the extent of the haze layers and the role of haze feedback in Pluto's atmosphere was completely unexpected. We outline here what was known about the photochemistry and potential for haze formation in Pluto's atmosphere before the New Horizons flyby of Pluto, how the observations from New Horizons have impacted our understanding of Pluto's atmospheric photochemistry and haze formation, feedback occurring due to the haze, and how these revelations impact our understanding of the evolution of Pluto's atmosphere.

### 1.1. Pre-New Horizons Picture of Pluto's Atmosphere

*1.1.1. Detection of Pluto's atmosphere.* Prior to the New Horizons flyby of Pluto, understanding of Pluto's atmosphere was limited to information gleaned from groundbased and Earth-orbiting telescopes. The first sign that Pluto may have an atmosphere was the detection of methane frost on the surface (*Cruikshank et al.,* 1976). This detection was made by making photometric observations of Pluto's reflectance in the infrared at 1.55 and 1.73 μm, which are the diagnostic wavelengths for methane and water ice absorption. By comparing these observations with laboratory measurements of the reflectance of methane ice in the wavelength range between 1 and 4 μm the authors determined that methane frost must be present on the surface. This detection meant that Pluto should have an atmosphere because methane ices could remain on the surface long term only if an atmosphere was present in vapor phase equilibrium with the surface (e.g., *Trafton,* 1980; *Fink et al.,* 1980; *Cruikshank and Silvaggio,* 1980). The first definitive detection of an atmosphere on Pluto was made with observations of an occultation of Pluto by a relatively bright (magnitude +12.8) star (*Brosch,* 1995). However, the first published detection of Pluto's atmosphere was during an occultation of Pluto by another relatively bright star that was observed by multiple telescopes in the southern hemisphere and the Kuiper Airborne Observatory (KAO). A stellar occultation occurs when Pluto passes between a star and the Earth. Measuring the reduction in brightness of the star as Pluto passes in front of it produces a "light curve" that provides information about the atmosphere and the solid body radius of Pluto. We illustrate in Fig. 1 the spatial coverage of Pluto's atmosphere for the telescopes observing the 1988 occultation (adapted from *Millis et*

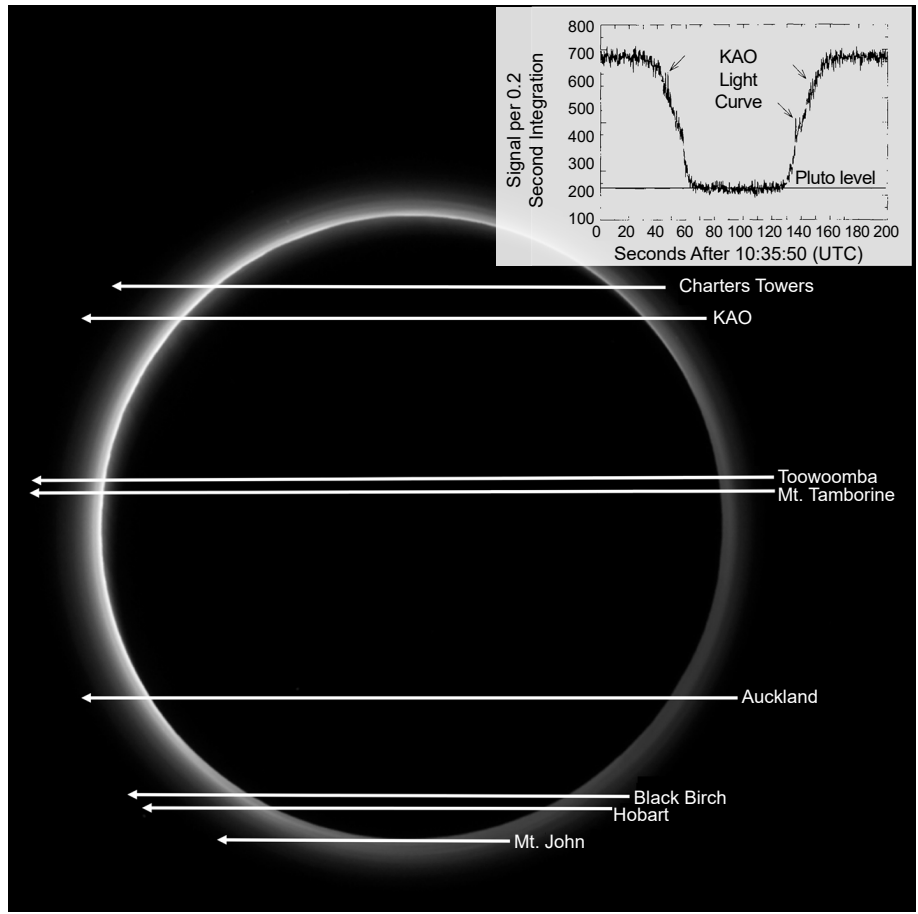

**Fig. 1.** Occultation path of the star P8 as observed from eight locations [adapted from *Millis et al.* (1993) to show paths over a New Horizons image of Pluto's global haze]. The light curve from Kuiper Airborne Observatory (KAO) in the inset shows time along the horizontal vs. signal level along the vertical axis (adapted from *Elliot et al.,* 1989). The gradual decline at the beginning of the occultation indicates detection of an isothermal atmosphere. Around the halfway point between the beginning of the occultation and complete occultation of the stellar signal at the "Pluto level" (solid line) there is an inflection point where the slope of the light curve increases. Arrows point to spikes that could indicate temperature variations in the atmosphere.

*al.,* 1993) and give an example light curve from this occultation (from *Elliot et al.,* 1989).

*1.1.2. Characteristics of Pluto's atmosphere.* The light curve provides a scale height, H, of the neutral atmosphere

$$H = \frac{kT}{m_a g}$$

where T is the atmospheric temperature, $m_a$ is the mean mass of the molecules in the atmosphere, k is the Boltzmann constant, and g is the force of gravity. The slope of the light curve is used to derive the scale height, obtaining information about atmospheric temperature and/or composition. Initial analysis of the occultation observation proposed that Pluto had an extended methane atmosphere as thick as Pluto's radius (*Hubbard et al.,* 1988). However, observations from KAO could be interpreted as an extensive atmosphere made up of either methane or nitrogen, depending on the assumed temperature (*Elliot*

*et al.,* 1989). *Yelle and Lunine* (1989) studied the energy balance of the atmosphere by simulating how energy input into and radiation from the atmosphere affects the temperature and concluded that a molecule heavier than methane must be present. The authors proposed that this molecule was either carbon monoxide (CO), molecular nitrogen ($N_2$), or argon (Ar), with a preference for CO because it was predicted to be the most dominant carbon-bearing molecule in the protosolar nebula (PSN). This was further demonstrated by thermal observations made between 1991 and 1993, which showed that the surface temperature of Pluto was too low to support a methane-dominated atmosphere, meaning that the atmosphere must be predominantly $N_2$ or CO (*Stern et al.,* 1993; *Jewitt,* 1994). Later detection of CO and $N_2$ ice on the surface showed that $N_2$ was the most abundant ice and that Pluto's atmosphere was likely to be predominantly $N_2$ with trace amounts of methane ($CH_4$) and CO (*Owen et al.,* 1993).

The light curve observations from KAO also showed a clear inflection point that could be interpreted as either a reversal in temperature variation as a function of altitude, also known as a temperature inversion, or a layer of haze near Pluto's surface, or some combination of both. This is illustrated in the inset to Fig. 1, where the slope of the light curve changes between the initial decline of the stellar signal and complete loss of the signal from the star behind Pluto (solid line in figure inset). Further analysis of the compiled observations found that the higher-quality observations confirmed either the presence of haze or a steep thermal gradient caused by a temperature inversion near the surface (*Millis et al.*, 1993). The authors suggested that haze provided a better explanation of the total quenching of light observed by KAO, but without two-color measurements a haze could not be proven. The next successful multichord stellar occultation of Pluto's atmosphere was in 2002 by the star P131.1. This occultation and subsequent ones extended observations to multiple wavelengths in the visible and infrared range, and showed a significant increase in surface pressure and further indications of a possible near-surface haze layer (*Elliot et al.*, 2003).

The potential for haze to form in Pluto's atmosphere was not surprising. Early studies of the photochemistry in Jupiter's (*Gladstone,* 1982; *Strobel,* 1983) and Titan's (*Yung et al.,* 1984) atmospheres discussed pathways for the production of haze initiated by photolysis of $CH_4$ that could logically be applied to Pluto (e.g., *Stansberry et al.,* 1989). The Voyager 2 flyby of Neptune detected haze below 30 km altitude in Triton's atmosphere (*Smith et al.,* 1989) and sparked comparative planetology studies between Triton and Pluto that led to improved models for photochemistry, the production of haze, and potential haze feedback in Pluto's atmosphere (e.g., *Krasnopolsky and Cruikshank,* 1995, 1999; *Summers et al.,* 1997).

Just as the results of the Voyager 2 flyby of Triton fed into an improved understanding of Pluto's atmosphere, the groundbreaking *in situ* and remote observations of Titan's atmosphere by the Cassini-Huygens mission (e.g., *Coates et al.,* 2007; *Crary et al.,* 2009; *Wahlund et al.,* 2009; *Brown et al.,* 2009) inspired valuable laboratory studies (e.g., *Dimitrov and Bar-Nun,* 2003; *Imanaka et al.,* 2004; *Trainer et al.,* 2013) and improved modeling for radiative transfer (e.g., *Yelle,* 1991; *Bampasidis et al.,* 2012) and photochemical processes (e.g., *Wilson and Atreya,* 2004; *Vuitton et al.,* 2006; *Lavvas et al.,* 2008; *De La Haye et al.,* 2008; *Krasnopolsky,* 2009; *Dobrijevic et al.,* 2016). These efforts resulted in significant advancements in understanding the photochemical processes relevant to Pluto's atmosphere leading into the New Horizons flyby.

The thermal structure of Pluto's atmosphere based on the occultation observations indicated the temperature was predicted to be high enough in the upper atmosphere to allow either hydrodynamic (e.g., *Stern and Trafton,* 1984; *Trafton,* 1990; *Krasnopolsky,* 1999; *Tian and Toon,* 2005; *Strobel,* 2008) or enhanced thermal escape of the high energy range of the velocity distribution function, known as Jeans escape (*Tucker et al.,* 2012), of the main constituent, $N_2$. The thermal structure plays an important role in pho-tochemical processes and in the long-term evolution of the atmosphere depending on whether escape or photochemistry have a major impact on the production and loss of volatiles in Pluto's atmosphere. A review of the composition and structure of Pluto's atmosphere is provided in the chapter in this volume by Summers et al., while Pluto's atmospheric escape is reviewed in the chapter by Strobel.

### 1.1.3. Seasonal variation of Pluto's atmosphere.
Pluto has a highly elliptical orbit with an orbital period of 248 years. The current (i.e., vs. long-term average) perihelion and aphelion distances are 29.66 and 49.31 astronomical units (AU) (distance of Earth from the Sun) respectively. This drastic difference in maximum and minimum distance from the Sun means that the flux of solar photons reaching Pluto is almost 3× greater at perihelion than at aphelion. This difference will result in variation of the surface pressure over a Pluto year, with debate as to whether or not the atmosphere would collapse at aphelion (*Stern and Trafton,* 1984; *Hanson and Paige,* 1996; *Young,* 2013; *Olkin et al.,* 2015). A review of the volatile and climate cycles on Pluto is provided in the chapter in this volume by Young et al., while the dynamics of Pluto's lower atmosphere are reviewed in Forget et al.

Pluto's atmosphere was first detected shortly before perihelion when Pluto was located at a distance of 29.76 AU from the Sun. Later occultations provided observations of the atmosphere from 2002 leading up to the New Horizons flyby on July 14, 2015, and covering a Pluto-Sun distance from 30.54 to 32.69 AU. Although the structure of the upper atmosphere did not appear to change significantly (*Elliot et al.,* 2007), the surface pressure increased until 2008 before beginning to decrease (see *Young,* 2013, and references therein), even though Pluto was moving away from the Sun during this entire time period.

Changes in surface pressure over Pluto's long year will influence the photochemical production of haze in several ways. First, a reduction in surface pressure decreases the amount of $N_2$ and $CH_4$ available for haze production. Reduced surface pressure may also indicate lower temperatures. This influences neutral reactions, which have a rate that depends on temperature, and increases condensation near the surface. Finally, the increasing distance from the Sun will decrease the flux of ultraviolet (UV) and extreme ultraviolet (EUV) photons available to initiate chemistry by photodissociation and photoionization of $N_2$ and $CH_4$, processes that processes that support Pluto's haze formation (see section 2). Although the 1988 occultations showed possible evidence of a haze layer near the surface, and occultations in 2003 showed indications of an extinction layer near the surface potentiallly attributable to aerosol condensation droplets (*Rannou and Durry,* 2009), other observations appeared to be a better fit with temperature inversion rather than haze (e.g., *Elliot et al.,* 2007; *Young et al.,* 2008). *Elliot et al.* (2007) proposed that haze may have formed around the 1988 occultations and later cleared. Leading up to the New Horizons flyby, one of the main questions related to photochemistry in Pluto's atmosphere was whether or not Pluto would have a near-surface haze layer similar to that observed at Triton (*Smith et al.,* 1989).

## 1.2. New Horizons Revelations

The New Horizons flyby provided several surprising discoveries about Pluto's atmosphere. The first was that the temperature in the upper atmosphere was significantly cooler than predicted, meaning that the atmosphere was being cooled by an unexpected process and was not escaping hydrodynamically (*Gladstone et al.,* 2016, hereafter *G2016*; *Young et al.,* 2018, hereafter *Y2018*). This discovery has major implications for the loss of volatiles and long-term evolution of the atmosphere, because escape had been predicted to be the dominant influence on how the atmosphere evolves (e.g., *Lunine et al.,* 1989; *Mandt et al.,* 2016). Additionally, photochemical reactions between neutrals are temperature-dependent, so a colder temperature profile influences photochemical production of complex organics that would lead to haze in Pluto's atmosphere.

The second surprise was the densities of hydrocarbons containing two carbon atoms ($C_2H_x$), or acetylene ($C_2H_2$), ethylene ($C_2H_4$), and ethane ($C_2H_6$), as a function of altitude. As shown in Fig. 2a, the densities of these species show an unexpected drop in density, or a density inversion, below 300–400 km. This density profile indicates that a loss process is at work removing these molecules from the atmosphere below 400 km. We explore this observation in detail in sections 2 and 4.

The final surprise from the New Horizons flyby was the extensive global layers of haze in the atmosphere (*Stern et al.,* 2015), illustrated in the full disk view of Pluto in Fig. 1 and in detail in Fig. 2b. We discuss these observations in further detail in section 3. Although the New Horizons observations confirmed the existence of haze in Pluto's atmosphere, the analysis of haze extinction profiles and the atmospheric temperature profile indicated that the inflection points recorded in the occultation light curves described above result from the strong temperature inversion in the lower atmosphere and not haze extinction (*Cheng et al.,* 2017).

## 1.3. Exploring the New Horizons Revelations

We discuss the implications of the New Horizons observations for understanding photochemistry and haze production in Pluto's atmosphere. Section 2 outlines the first steps in photochemistry, followed by section 3, which describes the haze observations and discusses what these observations indicate about molecular growth in Pluto's atmosphere. The presence of haze leads to feedback processes that are discussed in section 4. Then, in section 5 we discuss the implications of these new revelations for our understanding of how Pluto's atmosphere has evolved over time. We finish the chapter with a summary of what was learned, and a discussion of what work remains to be done through further analysis of the New Horizons datasets,

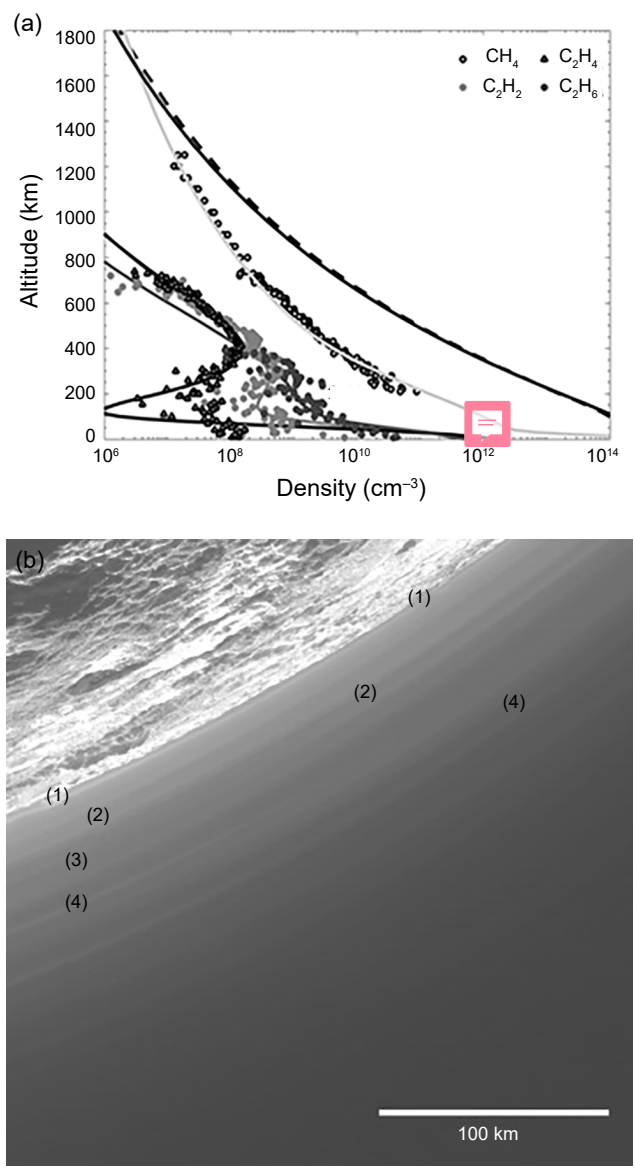

**Fig. 2.** **(a)** Densities of the main atmospheric species, molecular nitrogen ($N_2$) and methane ($CH_4$), and the $C_2H_x$ species, acetylene ($C_2H_2$), ethylene ($C_2H_4$), and ethane ($C_2H_6$), as a function of altitude showing a density inversion below 400 km. Model results are shown with the lines and suggest that haze particle hardening with aging is both necessary and sufficient to explain the inversion in the vertical profiles of $C_2$ hydrocarbons in Pluto's atmosphere [adapted from *Luspay-Kuti et al.* (2017) to use data from *Young et al.* (2018)]. **(b)** Several haze layers evaluated in *Cheng et al.* (2017).

future groundbased observations, and any potential future mission to the Pluto system.

## 2. PHOTOCHEMISTRY: FIRST STEPS

One-dimensional photochemical models provide a theoretical basis for observed expected atmospheric composition as a function of altitude by tracking the rate of production and loss for a given species through a system of chemical reactions. Model results vary relative to specified atmospheric conditions — including the vertical temperature profile, the incident solar flux at the top of the atmosphere, and the presumed upper and lower boundary conditions. Depending on the model application, the assumed atmospheric state may be defined relative to a specific latitude and local time, or relative to global average values. To retrieve the anticipated atmospheric composition profiles, photochemical models solve the one-dimensional continuity equation simultaneously for all specified atmospheric constituent species (*Banks and Kockarts,* 1973)

$$\frac{\partial n_i}{\partial t} + \frac{\partial \Phi_i}{\partial z} = P_i - L_i$$

where $n_i$, $\Phi_i$, $P_i$, $L_i$, are the species number density (in cm³), vertical diffusive flux (in cm² s⁻¹), and the chemical production and loss rates (cm³ s⁻¹), respectively, for each species i. Through this equation, the one-dimensional models trace the relative roles of the thermal structure, transport, and active chemical reactions between neutral and ion species created and/or destroyed by photolytic and kinetic processes on the overall vertical structure and composition of the atmosphere.

## 2.1. Pluto-Specific Photochemical Modeling

Prior to the New Horizons flyby of Pluto, the composition of Pluto's atmosphere was inferred from groundbased obser-vations of occultations as described above in section 1, and by applying photochemical models that had been validated for Triton using Voyager 2 observations (e.g., *Krasnopolsky and Cruikshank,* 1995, 1999; *Summers et al.,* 1997). The New Horizons flyby provided the first-ever altitude profiles of major constituents, N₂ and CH₄, as well as the minor constituent C₂ hydrocarbons (*G2016*).

Around the time of the New Horizons flyby, observations with the Atacama Large Millimeter Array (ALMA) provided an abundance measurement for CO, an altitude profile of hydrogen cyanide (HCN), and an upper limit for the detection of HC¹⁵N in Pluto's atmosphere (*Lellouch et al.,* 2017, hereafter *L2017*). The HCN observations are illustrated in Fig. 3. These observations provided new constraints for photochemical modeling of Pluto's atmosphere, inspiring three key modeling efforts reported to date in the literature. These models are (1) the Caltech/JPL KINETICS chemistry-transport model first developed by *Allen et al.* (1981) and updated for Pluto calculations by *Wong et al.* (2017) (hereafter *W2017*), which focuses solely on chemical processes occurring among the neutral gas species; (2) the coupled ion-neutral photochemical (INP) model developed initially for Titan by *De La Haye et al.* (2008) to evaluate the coupled ion and neutral chemistry and adapted to Pluto by *Luspay-Kuti et al.* (2017) (hereafter *LK2017*) and *Mandt et al.* (2017) (hereafter *M2017*); and (3) a coupled ion-neutral photochemical model developed by *Krasnopolsky* (2020) (hereafter *KR2020*) as an update to *Krasnopolsky and Cruikshank* (1999), which was developed prior to the New Horizons era and draws on knowledge gained from modeling Triton (*Krasnopolsky and Cruikshank,* 1995) and Titan (*Krasnopolsky,* 2009).

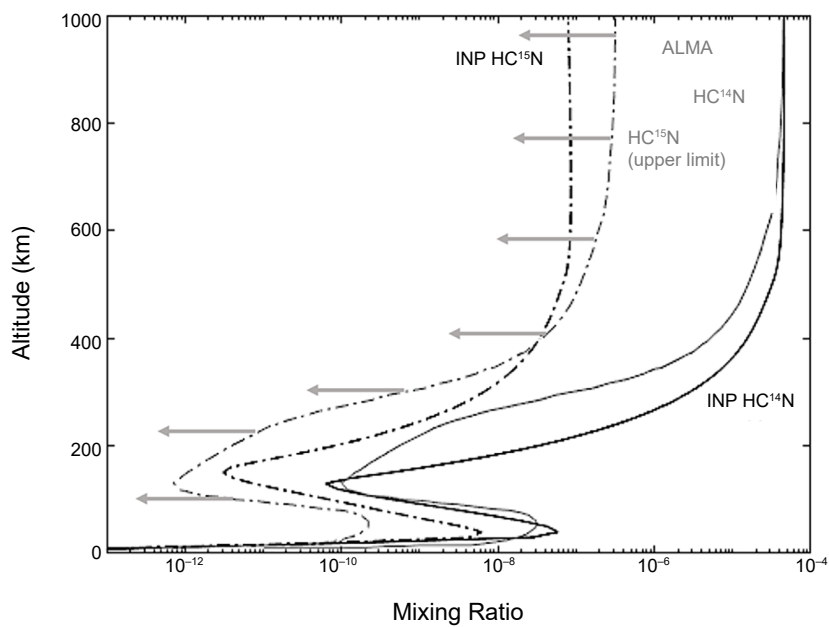

**Fig. 3.** Observed and modeled altitude profiles of hydrogen cyanide (HCN), comparing the measured HC¹⁴N profile and estimate HC¹⁵N upper limit (indicated with gray vertical arrows) from the Atacama Large Millimeter Array (ALMA) observations with the coupled ion-neutral photochemical model (INP) simulations of these species. Adapted from *Mandt et al.* (2017).

The comparative planetology leveraging modeling experience with Triton's and Titan's atmospheres has proven to be of high value, as understanding of each atmosphere is advanced by observations made by multiple missions. A model for Titan or Triton is converted to Pluto's conditions by adjusting the surface gravity, the vertical temperature profile, the heliocentric distance (which defines the incident solar flux at the top of the atmosphere), the anticipated incident energetic particles at the top of the atmosphere, the assumed surface pressure (which along with the surface T determines the sublimated $N_2$ and $CH_4$ density at the surface), and the assumed planet radius. Thin atmospheres can be simulated using a "plane-parallel" model that assumes all layers of the atmosphere are flat because cross sections show different levels of the atmosphere that appear flat. However, in the case of an atmosphere that extends far from the surface of the planet, the layers of the atmosphere curve along with the surface requiring spherical corrections. Pluto, Titan, and Triton all have extended atmospheres, so in all cases the model calculations are completed assuming a spherical atmosphere.

The vertical diffusive flux, $\Phi_i$ in equation (2), plays an important role in determining minor species densities as a function of altitude. It is calculated based on understanding of vertical eddy diffusion, which describes the turbulent mixing of the atmosphere, and molecular diffusion, which is specific to each species in the atmosphere. This flux varies as a function of temperature, thermal diffusion factor, and atmospheric scale height. Eddy diffusion dominates at lower altitudes where the total neutral density is greatest and maintains constant abundance with altitude for each species unless the constituent is significantly influenced by production and loss processes represented in the right side of equation (2). When molecular diffusion dominates, the species in the atmosphere will separate according to mass, allowing lighter species to become more abundant than heavier species with increasing distance from the surface. The altitude above which molecular diffusion becomes more dominant than eddy diffusion is called the homopause, and this is different for each species. This different homopause for each species occurs because each species has a unique value for molecular diffusion resulting in a different crossover altitude, or homopause, for each species (see example in Fig. 7 for $H_2$ and $CH_4$). Constraints for molecular diffusion are based on laboratory measurements made at temperatures much greater than those in Pluto's atmosphere and extrapolated to Pluto conditions with an uncertainty of less than 7% (*Plessis et al.,* 2015). Eddy diffusion is constrained by fitting an equation for the vertical flux, $\Phi_i$, to the altitude profile measured for a long-lived constituent.

In each model, boundary conditions are defined relative to the observed or expected atmospheric conditions. Each of the published models use the $CH_4$ mixing ratio derived from New Horizons (*G2016; Y2018*) as a fixed lower boundary condition. Although *LK2017* does not consider the incident anticipated $H_2O$ flux at the top of the atmosphere like the other two modeling efforts, the chemistry due to $H_2O$ is not expected to have a significant impact on the hydrocarbon or nitrile profiles because $H_2O$ primarily interacts with CO and its dissociative products (*Rodrigo and Lara,* 2002). *LK2017* begins with the surface $N_2$ gas density equivalent to the value expected for $N_2$ gas in vapor pressure equilibrium with the $N_2$ ice. The $N_2$ value near the surface is found in this model to increase with time in response to chemically driven $N_2$ production in the lower atmosphere. The other two modeling efforts assume the $N_2$ density at the surface is fixed. The extension of the $N_2$ vertical profile above the surface is a combination of the values observed by New Horizons at high altitude and the values calculated based on the temperature profile derived from the New Horizons stellar and radio occultation datasets (*G2016; Y2018*).

In each of the published models, solar EUV and UV radiation, interplanetary Lyman-α emission, and galactic cosmic rays initiate photochemical processes starting with photodissociation and photoionization. Although $N_2$ is the dominant gas in Pluto's atmosphere, $CH_4$ photodissociation is critical for the formation of more complex molecules that eventually lead to haze within the atmosphere. Each model also explores the role of gas species condensation as a function of altitude relative to New Horizons-era temperature profiles and laboratory measurements of the saturation vapor pressure curves, although there are some differences in the anticipated sensitivity of individual species to condensation and vertical mixing as a function of altitude as discussed in section 4.

Figures 4 and 5 illustrate how the photodissociation of $CH_4$ and $N_2$ initiates a chain of reactions leading to the production of hydrocarbons and nitriles [based on photochemical studies of Titan's atmosphere using the precursor model for *LK2017* [*De La Haye et al.,* 2008)]. These initial schemes of reactions are followed by further photodissociation and ionization processes, as well as neutral reactions and ion-neutral reactions eventually forming very large negatively charged macromolecules. It is important to note that the neutral reaction rates are temperature-dependent, so the chemistry at Titan and Pluto will have similar reactions but the rate of the reactions will differ as a result of the different temperatures in each atmosphere. The presence of CO and $CO_2$ in Titan's atmosphere has been shown to influence the composition of aerosols produced in laboratories simulating Titan-like conditions by allowing oxygen to be incorporated into the aerosols (*Trainer et al.,* 2004; *Hörst and Tolbert,* 2014). The presence of CO in Pluto's atmosphere should also allow for incorporation of oxygen into the haze as illustrated in Fig. 6 (adapted from *Trainer et al.,* 2004).

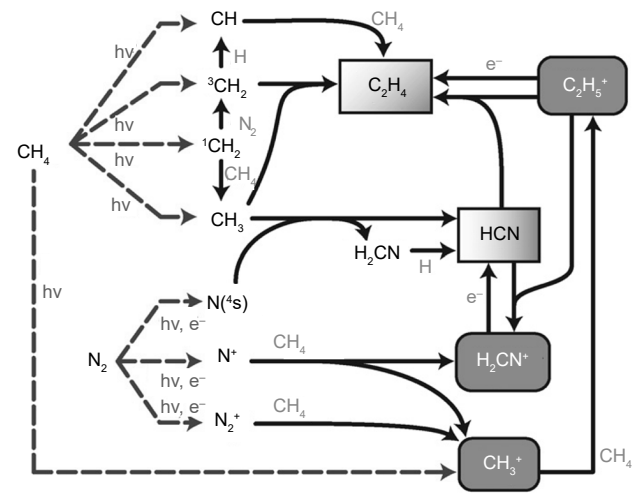

**Fig. 4.** Chain of reactions initiated by photodissociation of $CH_4$ and $N_2$ to produce the most basic nitrile, HCN, and one of the more basic hydrocarbons, $C_2H_4$. Ions that are produced in this scheme are shown in the dark gray squares and neutrals produced in the scheme are shown in the light gray squares. Adapted from chemical scheme I from *De La Haye et al.* (2008).

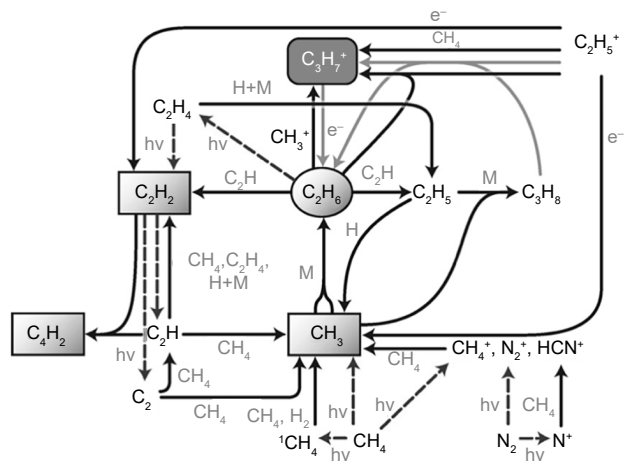

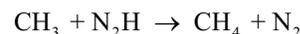

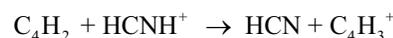

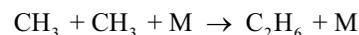

**Fig. 5.** Chain of reactions initiated by photodissociation of $CH_4$ and $N_2$ to produce more complex hydrocarbons. Ions that are produced in this scheme are shown in the dark gray squares and neutrals produced in the scheme are shown in the light gray squares. Adapted from chemical scheme II from *De La Haye et al.* (2008).

Although there are important differences between the three modeling studies described above, there is some agreement between them on what processes are involved in the production and loss of the species observed by New Horizons and ALMA in Pluto's atmosphere. We outline these processes in Table 1. The primary source of $N_2$ and $CH_4$ is the sublimation of ices on the surface. Methane is also produced by the photodissociation of larger hydrocarbons, and by bimolecular neutral reactions that transfer a hydrogen atom from one molecule to $CH_3$, an example of this type of reaction would be

$$CH_3 + N_2H \rightarrow CH_4 + N_2$$

The primary loss processes for both species are photodissociation and photoionization, escape from the top of the atmosphere. *LK2017* and *KR2020* both include ion chemistry and find that reactions between ions and molecules play a role in the production and loss of $CH_4$. Hydrogen cyanide is primarily produced through electron recombination of $HCNH^+$ and through the transfer of a proton from $HCNH^+$ to a neutral with a higher proton affinity. An example of this type of reaction is

$$C_4H_2 + HCNH^+ \rightarrow HCN + C_4H_3^+$$

The primary loss process for HCN is condensation onto aerosols and the surface as well as sticking to aerosols. The interaction with aerosols is discussed in greater detail in section 4. Finally, the $C_2$ hydrocarbons are produced primarily by photodissociation of larger hydrocarbons in the case of $C_2H_2$ and $C_2H_4$, where $C_2H_6$ is produced by three-body, or trimolecular, reactions. The main reaction producing $C_2H_6$ is

$$CH_3 + CH_3 + M \rightarrow C_2H_6 + M$$

where M is a molecule that facilitates the reaction but is not influenced by it. These types of reactions are only effective in the denser part of the atmosphere near the surface.

The primary loss process for the $C_2$ hydrocarbons is where the three modeling studies differ the most. *W2017* suggests that these species are lost by condensation onto aerosols while *LK2017* found that they could not condense and had to be lost by sticking to aerosols as described in section 4. Finally, *KR2020* suggests that these molecules flow to the surface where they condense as ices on the surface, forcing an inversion in the altitude profile. More work is needed to better understand loss processes for $C_2$ hydrocarbons in Pluto's atmosphere.

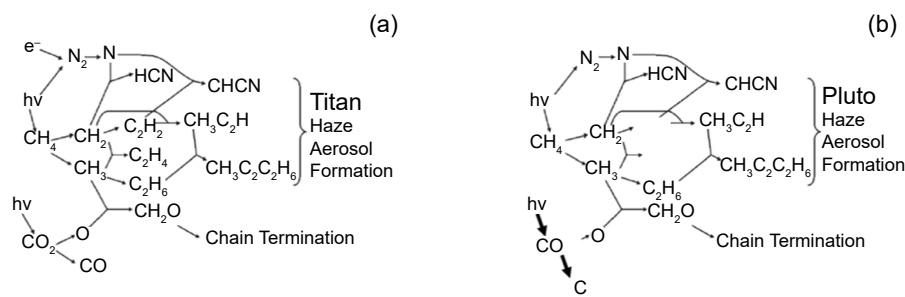

**Fig. 6.** Illustration of the incorporation of oxygen into the haze in **(a)** Titan's and **(b)** Pluto's atmosphere. Differences are in bold. Adapted from *Trainer et al.* (2004).

TABLE 1.  Main production and loss processes for species observed by New Horizons and ALMA in Pluto's atmosphere.

| | Production | Loss |
|---|---|---|
| $CH_4$ | Sublimation from surface | Escape |
| | H-transfer reactions | Photodissociation |
| | Ion-neutral reactions | Ion reactions |
| | Photodissociation of larger hydrocarbons | |
| $N_2$ | Sublimation from surface | Escape |
| | | Escape |
| HCN | Ion-neutral reactions | Sublimation to surface |
| | Electron recombination of $HCNH^+$ | Condensation onto aerosols |
| | | Adsorption onto aerosols |
| $C_2H_2$ | Photodissociation of larger hydrocarbons | Sublimation to surface |
| | | Adsorption onto aerosols |
| | | Ion-neutral reactions |
| $C_2H_4$ | Photodissociation of larger hydrocarbons | Sublimation to surface |
| | | Adsorption onto aerosols |
| | | Ion-neutral reactions |
| | | Photodissociation |
| $C_2H_6$ | Trimolecular neutral reactions | Sublimation to surface |
| | | Adsorption onto aerosols |
| | | Ion-neutral reactions |
| | | Photodissociation |
| | | Sublimation to surface |

## 2.2.  Vertical Dynamics Represented by the Eddy Diffusion Coefficient

Methane has a relatively long chemical lifetime in Pluto's atmosphere and is sensitive to transport processes. This makes the altitude profile observed by New Horizons the most reasonable constraint available for inferring vertical eddy diffusion, or $K_{zz}$, in Pluto's atmosphere. Unfortunately, as illustrated in Fig. 7, the published literature presents $K_{zz}$ values as low as $10^3$ (*W2017*; *Y2018*) and $10^4$ (*KR2020*), and as high as $10^6$ cm$^2$ s$^{-1}$ (*G2016*; *L2017*; *LK2017*). This broad range of inferred diffusion rates is difficult to evaluate in detail due to limited information provided in the published literature. It is possible that the differences are linked to how the boundary conditions assumed in each model are applied to the diffusion calculation.

*W2017* reports the lowest values using the KINETICS 1-D PCM, which calculates transport independent of any ion or neutral chemistry or any consideration of $CH_4$ escape rates. KINETICS uses the classical formulation of the diffusion equation (*Banks and Kockarts,* 1973) like *G2016*, but does not provide sufficient explanation as to why their result is so much lower than the initial estimate reported in *G2016*. This makes it difficult to determine why the result changed and is so different from the values derived by *L2017* and *LK2017*, which use a slightly modified one-dimensional diffusion equation. *Y2018* also report a very low value using the *Strobel et al.* (2009) one-dimensional transport model, which includes escape and requires an upward flux of methane from the surface to balance escape at the top of the atmosphere. On the other hand, *KR2020* derives a value that is an order of magnitude larger by fitting not only to $CH_4$, but also to the $C_2$ hydrocarbons assuming a downward flux of the hydrocarbons due to condensation at the lowest altitudes. Finally, the calculations completed by *LK2017* apply a fixed lower boundary mixing ratio for $CH_4$ relative to $N_2$ and allow thermal escape at the top of the atmosphere.

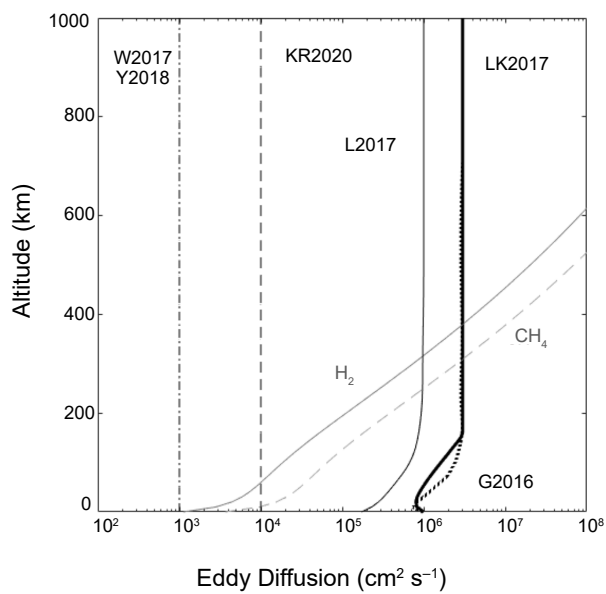

**Fig. 7.** Reported eddy diffusion coefficients in the published literature for the post–New Horizons analysis of Pluto's atmosphere compared with molecular diffusion for $H_2$ and $CH_4$. *W2017* and *Y2018* have the lowest reported eddy diffusion, while *G2016*, *L2017*, and *LK2017* have the highest. *KR2020* has an intermediate value.

There are slight differences in the mixing ratio for methane at the surface in each of the models. Although there is some sensitivity in the inferred $K_{zz}$ values to the conditions in the surface boundary layer, it is not sufficient to cause variation in the inferred $K_{zz}$ value by ~2 to 3 orders of magnitude. There is also a small difference in the formulation for the diffusive flux equation in *W2017* compared to *LK2017*. The formulation used in *W2017* is the classical equation (*Banks and Kockarts,* 1973), while INP uses a formulation modified for a two-component atmosphere (*De La Haye,* 2005; *De La Haye et al.,* 2007). The primary difference between these formulations is that the thermal diffusion coefficient is multiplied by a factor of $[(n-n_i)/n]$, where n is the total number density and $n_i$ is the number density of species i. In the case where the $CH_4$ and $N_2$ atmospheric densities are comparable, the thermal diffusion coefficient would be multiplied by a factor of ~0.5. However, this is only relevant above 800 *km* in Pluto's atmosphere. At all lower altitudes, the $CH_4$ and $N_2$ line-of-sight abundance profiles differ by 2 orders of magnitude, so the $(n-n_i)/n$ term is ~1 at these altitudes and there is no real difference in the classical and modified diffusive flux calculations. Therefore, this difference in the diffusive flux equation cannot explain the difference in results between *W2017* and *LK2017*. Aside from the formulation of the diffusion equation the main differences between the models is in the handling of the methane lower boundary, and whether $CH_4$ loss to photochemical processes (photoloysis, neutral, and ion-neutral reactions) is included in determining the eddy diffusion profile. These two factors are likely the main source of difference in the results in these models.

### 2.3. Including Gas-Aerosol interaction Proxies Within the Pluto Photochemical Models

A complete description of the processes impacting Pluto's atmospheric structure would require that the continuity equation is defined as a function of losses and gains occurring as a result of solar driven chemistry, kinetic chemistry, polymerization (leading to aggregate macromolecule formation and sedimentation), condensation (i.e., gas species phase changes), and gas-involatile interactions (which may proceed by two different processes: either condensation onto condensation nuclei, or bonding via coagulation to and coalescence into a haze particle), diffusion, escape, sublimation, and condensate precipitation. Interactions between gas and hazes are discussed in greater detail in section 4.

### 3. MOLECULAR GROWTH LEADING TO AEROSOL AND HAZE FORMATION

Haze formation in Pluto's atmosphere is the natural endpoint of the photodissociation of methane and nitrogen and the production of more complex species. Although the molecules detected by New Horizons are limited to $N_2$, $CH_4$, and $C_2$ hydrocarbons, with HCN and CO detections from ALMA (section 2), photochemical models of Pluto's atmosphere predict a panoply of higher-order hydrocarbons and nitriles (e.g., *Krasnopolsky and Cruikshank,* 1999; *W2017*) that could act as the precursors to the global haze.

### 3.1. New Horizons Observations of Pluto's Global Haze

Pluto's haze was observed by several instruments onboard New Horizons, spanning more than an order of magnitude in wavelength and multiple scattering angles around the planet. *Cheng et al.* (2017) summarized the findings from the Long Range Reconnaissance Imager (LORRI) instrument, which observed at wavelengths between 350 and 850 nm and scattering angles from 20° (nearly backscattering) to 169° (nearly forwardscattering). From full-disk images, these workers found that the haze extends completely around the limb of Pluto with greater haze brightness in Pluto's northern hemisphere than its equatorial latitudes and southern hemisphere. The scale height of the haze brightness was ~50 km and decreased as the altitude increased. The most striking feature of the haze is the ~20 concentric layers (Fig. 2b) that merge, separate, and appear or disappear when traced around the limb; their average thickness is on the order of kilometers, but spatially variable (*G2016*).

In an effort to explain the greater haze brightness in the northern hemisphere rather than the subsolar point during the occultation observations, *Bertrand and Forget* (2017) coupled a three-dimensional global climate model (GCM) for Pluto with a parameterization for haze production. This GCM tracks the sublimation, transport, and condensation of nitrogen and methane on the surface and was updated to evaluate the production and transport of haze up to altitudes of 600 km. The authors were able to demonstrate that the methane loss rates would be highest at northern latitudes around 250 km in altitude leading to the greatest haze production at this location rather than at the subsolar point. This occurs because the flux of photons at Lyman-α wavelengths was greatest at these latitudes during the New Horizons flyby.

The LORRI haze observations cannot be explained by a single haze particle shape (*Cheng et al.,* 2017). Near the surface, the haze is strongly forwardscattering while also exhibiting a significant backscattering lobe, similar to comet dust (e.g., *Ney and Merrill,* 1976; *Kolokolova et al.,* 2004), which is made up of spherical particles that contain ices. However, at 45 km above the surface, the backscattering lobe is missing while the forwardscattering remains intense (*Cheng et al.,* 2017), signaling a transition in shape from spherical to fractal aggregates — fluffy conglomerations of smaller particles with significant void space (e.g., *West and Smith,* 1991; *Lavvas et al.,* 2010). The existence of aggregates is also supported by the Multispectral Visible Imaging Camera (MVIC), which observed the haze in blue (400–550 nm) and red (540–700 nm) wavelengths. *G2016* found that the haze was brighter in the blue channel, suggesting scattering by small (~10 nm) particles, which contrasts with the intense forwardscattering typical of larger (>100 nm) particles. This seeming discrepancy can be resolved by invoking aggregates, since they are made of small particles but their bulk radius can be much larger (*G2016*).

*Y2018* summarized the findings of the Alice ultraviolet spectrograph, which observed solar occultations of Pluto's atmosphere from 52 to 187 nm. At these shorter wavelengths, the haze is much more opaque, and as such was detectable up to an altitude of 350 km. The scale height of the haze in the UV is also larger, at ~70 km. Importantly, like the forward- and backscattering LORRI observations, the Alice and LORRI data also cannot be explained by a single haze particle shape: Spherical particles that fit the LORRI data at all phase angles underestimate the UV extinction seen by Alice, while aggregate particles that fit the UV extinction seen by Alice and the forwardscattering data from LORRI underestimate the backscattering observations (*Cheng et al.,* 2017). This suggests that the shape — and thus the formation process — of near-surface haze particles are more complex than assumptions of only spheres or only aggregates and that the reality is likely a combination of both.

### 3.2. Lessons from Titan: Haze Precursors and the Formation of Monomers

We show in Fig. 8 a comparison of the temperature vs. pressure in the atmosphere of Pluto compared to that of Titan, Saturn's largest moon, and Triton, Neptune's largest moon. Pluto's atmospheric composition is similar to both atmospheres, as all three are composed primarily of $N_2$, with CH$_4$ as the second most abundant constituent. The Cassini-Huygens mission's exploration of Titan revealed a complex web of neutral and ion reactions beginning above an altitude of 1000 km. There, the photodissociation and ionization of N$_2$ and CH$_4$ due to EUV and FUV photons and energetic particles from Saturn's magnetosphere create an ionosphere, which is also fed by metallic ions from the ablation of meteorites. The detection of aerosols at altitudes as high as the ionosphere (*Liang et al.,* 2007) and large ions in the upper atmosphere (e.g., *Waite et al.,* 2007; *Crary et al.,* 2009; *Coates et al.,* 2009) suggests that haze formation begins in the ionosphere, although the production pathway is not fully understood.

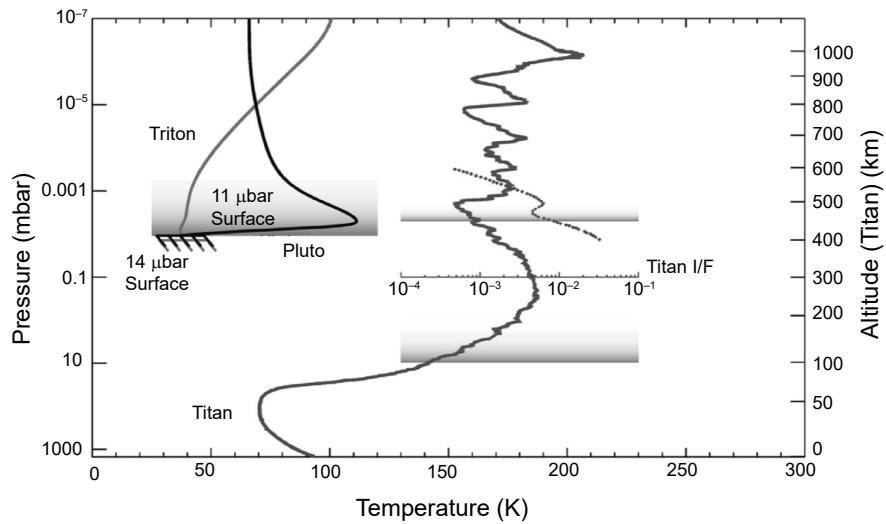

**Fig. 8.** Comparison of the atmospheric temperature and pressure profiles for Titan based on Cassini Huygens probe observations, Triton from Voyager 2, and Pluto from New Horizons. Also shown are the relevant altitudes for these pressures in Titan's atmosphere and the Titan I/F for the altitude range ~400–600 km. The shaded regions indicate the locations of the haze layers in Titan's and Pluto's atmospheres. Because Triton's haze was only observed up to ~30 km, which is very close to the surface, the Triton haze layer is not shown. Adapted from Fig. 22 of *Cheng et al.* (2017).

One proposed production pathway consists of ion and neutral chemical reactions rapidly producing haze precursor molecules such as benzene (C$_6$H$_6$) and heavier polycyclic aromatic hydrocarbons (PAHs) (*Wilson and Atreya,* 2004, *Trainer et al.,* 2013, *Yoon et al.,* 2014), followed by collision and sticking of the PAHs that then create small spherical particles with radii of ~0.1–0.5 nm (*Lavvas et al.,* 2011). These particles become negatively charged due to the absorption of free electrons in the ionosphere. The negative charge attracts positive ions, leading to rapid growth of particles to radii >1 nm (*Lavvas et al.,* 2013). Finally, neutral heterogeneous reactions on the surfaces of these particles result in the formation of spherical particles with radii of several nanometers (*Lavvas et al.,* 2011), which can then collide and stick to form the observed haze particles. A review of Titan's atmospheric chemistry and haze formation processes can be found in *Hörst* (2017).

There also exist, however, several key differences between Pluto and Titan that call into question the applicability of haze formation processes inferred in Titan's atmosphere to Pluto. The most striking difference is the energetics of haze formation. New Horizons failed to detect an ionosphere at Pluto and constrained the ion density to be <1000 e$^-$ cm$^{-3}$ (*Hinson et al.,* 2018), 5× less than that of Titan (*Gladstone and Young,* 2019). This may not be surprising, since Pluto receives less than a tenth of the solar UV flux received by Titan and there is no giant planet magnetosphere to source energetic particles. On the other hand, Lyman-α radiation scattered from H atoms in the local interstellar medium could be a significant driver of photochemistry (*W2017*), and deposition of ions stemming from the ablation of water ice and silicate interplanetary dust particles could feed a low-density ionosphere (*Poppe and Horányi,* 2018). These differences in upper atmosphere energetics could lead to variations in haze formation pathways and composition. In addition, the CO mixing ratio of Pluto's atmosphere [~0.05% (*Gladstone and Young,* 2019)] is ~10× larger than that of Titan. To explore this, laboratory investigations (e.g., *Hörst and Tolbert,* 2014; *Fleury et al.,* 2014, *He et al.,* 2017) have shown that changing CO mixing ratios in a N$_2$-CH$_4$ gas mixture exposed to high-energy sources could lead to dramatically different haze production rates, particle sizes, and composition.

## 3.3. Aggregation and Transport of Aerosols

The particle size and number density of aerosols in planetary atmospheres are controlled by the balance between microphysical and transport processes. Microphysical processes refer to those that directly impact the size of particles, including growth by uptake of volatile vapors (condensation) and the adsorption of gases (sticking), or by colliding and sticking with other aerosol particles (coagulation and coalescence), as well as loss due to evaporation or fragmentation. Transport processes include sedimentation of aerosols under the action of gravity, as well as the diffusion of particles due to Brownian motion (Brownian diffusion) and large-scale atmospheric dynamics (e.g., eddy diffusion and advection).

The low temperatures and densities of Pluto's atmosphere means that only a subset of microphysical and transport processes need to be considered. In particular, sedimentation of aerosols due to gravity dominates over diffusion and dynamics for most of Pluto's atmosphere (*Gao et al.,* 2017, *Bertrand and Forget,* 2017). As a result, small spherical particles growing through ion and neutral chemistry in the putative Pluto ionosphere can only grow to a certain size before their sedimentation timescale becomes shorter than their growth timescale, and they begin their journey toward the surface. This size should be on the order of a few to 10 nm, as inferred from MVIC observations (section 3.1), similar to analogous particles in Titan's atmosphere (section 3.2). Furthermore, because the mean free path of aerosols is large compared to their size — their Knudsen numbers are large — the sedimentation velocity becomes inversely proportional to the atmospheric density and the square root of the temperature, resulting in decreasing fall speeds toward Pluto's surface (*Gao et al.,* 2017).

Nascent haze particles can collide with each other during their descent and stick together with weak intermolecular adhesion forces in the process of coagulation (*Dimitrov and Bar-Nun,* 1999). For sufficiently rigid particles, sticking may occur at only one point between them, forming fractal aggregates with the constituent particles termed "monomers." The rate of coagulation is proportional to the square of the particle number density and the coagulation kernel, which is a rate coefficient used to calculate coagulation probabilities that takes into account the thermal velocity (Brownian coagulation) and relative fall velocities (gravitational collection) of the colliding aerosol particles. At large Knudsen numbers, the Brownian coagulation kernel becomes proportional to the square root of temperature (*Lavvas et al.,* 2010), suggesting increasing coagulation rates with decreasing altitude between a few hundred to tens of kilometers above Pluto's surface, where the temperature also increases with decreasing altitude. In contrast, laboratory studies conducted in preparation for the Cassini-Huygens mission observations at Titan found that aerosols produced in the laboratory hardened as they grew in size and "aged," making them less sticky (*Dimitrov and Bar-Nun,* 2002). This effect would lead to a decrease in coagulation rates with altitude.

*Gao et al.* (2017) computed the haze particle number density and size distribution in Pluto's atmosphere using an aerosol physics model and compared their results to haze extinction observations from the Alice ultraviolet spectrograph (section 3.1). They assumed that the haze production rate was equal to the methane photolysis rate computed by the photochemical model of *W2017*. *Gao et al.* (2017) found that, above 200 km, haze particles remained small, either as individual monomers or small aggregates, due to rapid sedimentation and low coagulation rates (Fig. 9). Below 200 km, decreasing fall speeds and increasing coagulation kernels result in the growth of aggregates through coagulation, ultimately forming aggregates with radii ~100–200 nm near the surface, consistent with LORRI data (*G2016*; *Cheng et al.*, 2017). *Gao et al.* (2017) also simulated an alternative case where all haze particles were spherical, but found that they underestimated the particle size and ultraviolet extinction.

The ultraviolet extinction predicted by the aggregate model of *Gao et al.* (2017) increases with decreasing altitude faster than that observed by the New Horizons Alice UV spectrograph, as illustrated in Fig. 10. Curiously, this can be explained if the monomers within the fractal aggregates grew with decreasing altitude. While 5–10-nm monomers can reproduce the data above 150 km, 25-nm monomers are needed below 100 km. Monomers can grow through coalescence: If there is enough excess energy available after coagulation, then the intermolecular adhesion forces can be replaced by stronger, cohesive intermolecular bonds (*Dimitrov and Bar-Nun*, 1999). At this stage, the monomers are partially fused together while still keeping their chemical structures. If there is still considerable energy left in the system then coalescence will continue and the particles will fuse together, forming spheres. Another way for monomers to grow is through condensation and/or adsorption of atmospheric gases onto the aggregate, which will be elaborated on in section 4. The growth and possible coalescence of monomers with time and decreasing altitude may provide an explanation for the discrepancy between the multi-wavelength, multi-phase angle observations of Alice and LORRI on New Horizons (section 3.1).

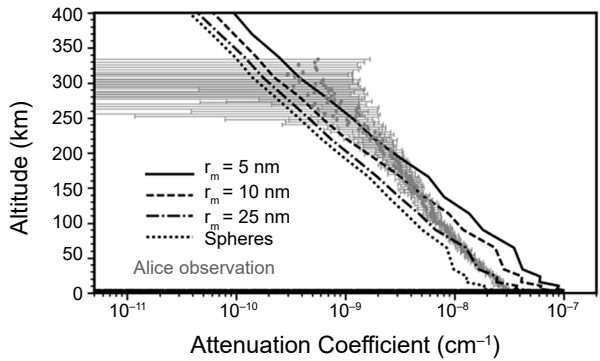

**Fig. 10.** The New Horizons Alice observed (points) and model attenuation coefficient (curves) of Pluto's haze. Models with monomer radii of 5 nm (solid), 10 nm (dashed), and 25 nm (dash-dot) are presented along with the attenuation coefficient of spherical haze particles. All haze models have the same production rate. Updated from *Gao et al.* (2017).

## 4. HAZE FEEDBACK PROCESSES IN PLUTO'S ATMOSPHERE

As described in the previous sections, there are many similarities in photochemistry and haze formation in Pluto's and Titan's atmospheres that allow application of lessons learned from Titan modeling to Pluto. Of particular interest is how the formation and growth of haze feeds back into atmospheric processes. Observations made during the New Horizons flyby provide important constraints on these feedback processes. Studies of Titan's atmosphere found that the haze particles play an important role in shaping the vertical abundance profiles of hydrocarbon and nitrile species when the molecules condense onto and irreversibly stick to the haze particles (*Willacy et al.*, 2016). This influence of haze in Titan's atmosphere raises the question of whether haze particles play a similar role in Pluto's atmosphere.

An early indication for possible haze feedback in Pluto's atmosphere was the shape of the vertical $C_2$ hydrocarbon profiles retrieved from New Horizons. Instead of continuously increasing toward the surface, as would be expected if photochemistry were the only process affecting their profiles, the densities of $C_2H_2$, $C_2H_4$ and $C_2H_6$ showed an inversion between 100 and 500 km [illustrated in Figs. 2 and 11 (*G2016*; *W2017*; *LK2017*; *Y2018*)]. The observed inversion in the altitude range where haze particles were also observed suggests that haze-molecule interactions dominate hydrocarbon loss in this altitude range. Furthermore, the altitude profile of HCN determined with the ALMA observations (*L2017*) also shows haze-particle interaction because HCN is supersaturated, or has partial pressures that are orders of magnitude larger than the saturation vapor pressure, above

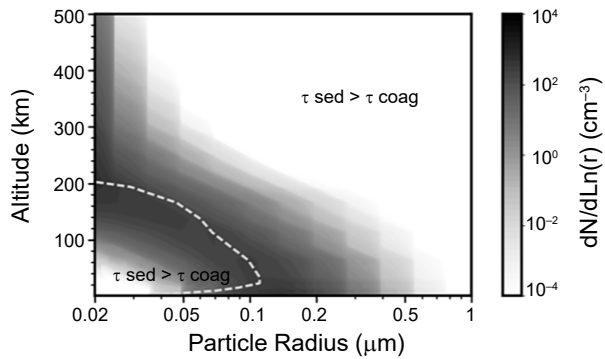

**Fig. 9.** Particle number density as a function of altitude and particle radius, assuming a fractal aggregate shape for the particles. The peak of the particle size distribution can be traced by balancing the sedimentation timescale with the coagulation timescale (dashed line), indicating that the particles can grow until sedimentation transports them to lower altitudes. Updated using the model from *Gao et al.* (2017) described in section 3.3.

150 km and drops to subsaturated, or partial pressures below the saturation vapor pressure, around ~50 km.

## 4.1. Condensation onto Haze Particles

Whether a species in Pluto's atmosphere will condense onto haze particles or not depends on the collision rate between the molecule and the particles, the availability of condensation nuclei, and the molecule's partial and saturated vapor densities. Condensation requires that the density of an atmospheric constituent be greater than its saturation vapor density. The saturation vapor density of a species is derived from its saturation vapor pressure. The saturation vapor pressures of the species observed in Pluto's atmosphere have been calculated at their vapor-solid equilibrium based on limited laboratory data. However, the temperature range over which these laboratory measurements were performed does not extend down to the lowest temperatures observed in Pluto's atmosphere. Thus, extrapolation to low temperatures is often needed based on expressions determined from laboratory experiments at higher temperatures.

Figure 12 shows the calculated saturation vapor pressures and densities for species observed in Pluto's atmosphere using equilibrium expressions available in the literature. When comparing the saturation vapor densities to the atmospheric densities derived from New Horizons measurements, it is apparent that for the most part, the hydrocarbon densities are significantly lower than their saturation vapor densities through most of the atmosphere. The HCN altitude profile was observed by ALMA and shows that HCN is supersaturated above 150 km (*L2017*). *Rannou and West* (2018) evaluated condensation processes in Pluto's atmosphere using a model based on the classical laws of cloud nucleation, or the formation of clouds by the phase transition from supersaturated air into droplets. Nucleation can occur either by heterogenous nucleation, which is nucleation onto a surface, or through homogeneous nucleation, which occurs away from a surface. Heterogenous nucleation is more common than homogeneous nucleation, and in the case of Pluto's atmosphere would take place on the surfaces of photochemically produced haze particles. *Rannou and West* (2018) found that homogeneous nucleation is unlikely in Pluto's atmosphere, requiring the presence of aerosols as nucleation sites for heterogenous nucleation, thus allowing HCN to reach high supersaturation above the layers of haze.

Based on this, HCN is able to condense throughout the atmosphere of Pluto, but the $C_2$ hydrocarbons cannot condense except for within a thin, near-surface layer. The one exception is $C_2H_2$, which may condense between 300 and 500 km if the saturation vapor pressure expression from *Moses et al.* (1992) is used (*LK2017*). This implies that condensation onto haze particles does not affect the $C_2H_4$ and $C_2H_6$ vertical density profiles. Furthermore, the photochemical model of *LK2017* suggests that condensation onto haze particles is negligible even for $C_2H_2$, which may become saturated over a narrow altitude range. Thus, *LK2017* conclude that condensation onto the haze particles is not responsible for the observed density inversion in the hydrocarbon density profiles.

These results are in contradiction to the model results of *W2017*, who conclude that condensation is the major loss process for the hydrocarbons, and that in order to reproduce the inversion of $C_2H_4$, its saturation vapor pressure must be much lower than what would result from extrapolation of laboratory data to the low temperatures of Pluto's atmosphere. *W2017* suggest that the saturation vapor pressure of $C_2H_4$ must be the same as that of $C_2H_6$ at the low temperatures of Pluto, although they provide no physical explanation as to why this would be the case. The currently available published laboratory data, and empirical and thermodynamic relations show no indication for such behavior of the $C_2H_4$ vapor pressure.

The atmospheric profile of HCN appears to be strongly influenced by condensation onto the haze particles (*M2017*; *W2017*; *L2017*). Based on Fig. 11, any HCN densities greater than $10^8$ cm$^{-3}$ are subject to condensation, given sufficient availability of condensation nuclei. The efficiency with which HCN molecules condense onto haze particles is roughly a factor of 2 smaller determined by *M2017*, who used the photochemical model described in *LK2017*, than that found by *W2017*. Although it is important to note that while the treatment of condensation is stated to be the same based on the approach of *Willacy et al.* (2016) in both photochemical models, the equation given by *W2017* does not include a term for the saturation vapor pressure, and more closely resembles the equation for incorporation into aerosols (see below in section 4.2). Because of this inconsistency, a direct comparison of the results of *W2017* to the results of *LK2017* and *M2017* is challenging.

## 4.2. Incorporation into Haze Particles

Atmospheric molecules may irreversibly stick to and become incorporated into haze particles, leading to their permanent loss (*Willacy et al.,* 2016). As such, this haze feedback process has the potential to influence the altitude profiles of species in Pluto's atmosphere. This incorporation of molecules into haze particles is independent of the saturation vapor density. The loss rate due to incorporation into haze particles depends on the mean surface area of haze particles per volume as a function of altitude, the sticking efficiency of the molecules, the thermal velocity of the given molecule, and the atmospheric density of the molecule at a given altitude (*Willacy et al.,* 2016).

The formation and distribution of haze particles in Pluto's atmosphere were recently modeled based on New Horizons measurements by *Gao et al.* (2017) and show that the haze particles' surface area increases with decreasing altitude (Fig. 12a). Using this altitude profile of the haze particles, recent photochemical models quantified the loss of the detected molecules in Pluto's atmosphere due to their irreversible sticking to haze particles. As discussed in detail section 3.3, haze particles are sticky when they first form, and become less sticky as they age due to harden-ing. The sticking efficiency of a given molecule factors in both the ability of a given molecule to stick to the haze particles and the stickiness of the haze particles themselves according to their hardening/aging. It is important to note when comparing Pluto and Titan that Pluto's atmosphere receives lower UV radiation and experiences significantly lower temperatures than Titan's atmosphere. This leads to lower collision rates between the molecules and the haze particles and lower-energy deposition rates compared to Titan. Thus, we expect the hardening and aging of haze particles to be slower at Pluto, which could potentially increase the rate of hydrocarbon and nitrile incorporation into Pluto's haze (*LK2017*).

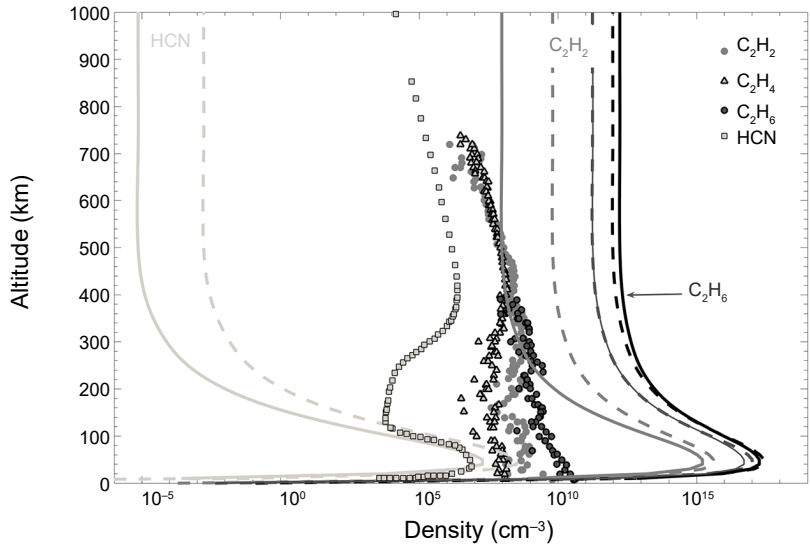

**Fig. 11.** Comparison of the saturation vapor densities to the New Horizons C$_2$H$_2$ (medium gray), C$_2$H$_4$ (dark gray), and C$_2$H$_6$ (black) densities and the ALMA HCN (light gray) densities. The solid lines are saturation vapor densities as listed in Table 2 of *Willacy et al.* (2016), while the dashed lines are saturation vapor densities from *Fray and Schmitt* (2009). This comparison demonstrates that HCN is supersaturated above 150 km and subsaturated around 50 km, while the C$_2$ hydrocarbons are subsaturated in almost all altitude ranges. Modified from *LK2017* and *M2017*.

The sticking efficiencies of the C$_2$ hydrocarbons and HCN were empirically determined by fitting the densities observed by New Horizons and ALMA with modeled densities. The sticking efficiencies show that the stickiness of haze particles in Pluto's atmosphere is inversely proportional to their mean surface area (Fig. 12b). Thus, the ability of molecules to stick to them also decreases with increasing particle size (*LK2017*; *M2017*). This means that the larger that the aerosol surfaces available to collide with and stick to are, the fewer hydrocarbons and HCN are actually able to stick to them. The decreasing stickiness of haze particles with increasing haze particle size (which increases with decreasing altitude) in Pluto's atmosphere is demonstrated in Fig. 12. With the sticking efficiencies and haze stickiness empirically derived for Pluto's hydrocarbons, the observed inversion in their vertical density profiles by New Horizons was successfully reproduced by INP [Fig. 2a (*LK2017*)]. The fact that the condition of molecules sticking to the haze is necessary to reproduce the density inversion in Pluto's atmosphere indicates that haze particles do indeed harden, or become less sticky as they age, similar to what was predicted by *Dimitrov and Bar-Nun* (2002) for Titan's aerosols.

### 4.3. Atmospheric Thermal Balance

In addition to the inversion of the vertical density profiles of the C$_2$ hydrocarbons, another issue raised by New Horizons measurements was the significantly lower atmospheric temperature than predicted by pre-New Horizons theory (*G2016*; *Zhu et al.,* 2014), as shown by the comparison of temperature profiles in Fig. 13. The colder-than-expected temperatures suggest some kind of a cooling mechanism in Pluto's atmosphere. While typically atmospheric gases determine the radiative energy balance of planetary atmospheres, they were found to be insufficient to cause the observed low temperatures in Pluto's atmosphere (*Strobel and Zhu,* 2017). However, hazes are another source of heating and cooling in planetary atmospheres. *Zhang et al.* (2017) assessed the radiative effects of hazes in Pluto's atmosphere by a multiscattering radiative transfer model and found that the heating and cooling rates caused by Pluto's hazes are 1 to 2 orders of magnitude larger than the gas cooling rates. Thus, Pluto's atmosphere is unique among planetary atmospheres in the sense that the total radiative energy budget is driven by haze particles instead of gas molecules. In other words, Pluto's atmospheric gas temperature appears to be controlled by the haze particles instead of the gases themselves below an altitude of about 700 km. Above 700 km, heat conduction drives the temperature, according to *Zhang et al.* (2017).

A caveat for this thermal balance model is that it must rely on estimated optical properties of the haze, called optical constants. Currently no observational constraints on the composition of these haze particles exist. The lack of detailed knowledge in the haze composition may lead to large uncertainties in the modeled heating and cooling rates. If hazes are indeed the drivers of atmospheric cooling on Pluto, then they are expected to radiate large amounts of heat into space. In that case, Pluto would appear several orders of magnitude brighter in the mid-infrared wavelength range than previously thought. The James Webb Space Telescope (JWST), expected to be launched in 2021, should be able to detect this predicted brightening (*Zhang et al.,* 2017). Because JWST could directly detect the radiative effects of haze, future observations would help to differentiate be-tween the effects of haze and gas molecules such as water vapor. This would be a giant leap toward settling the issue of Pluto's atmospheric temperature.

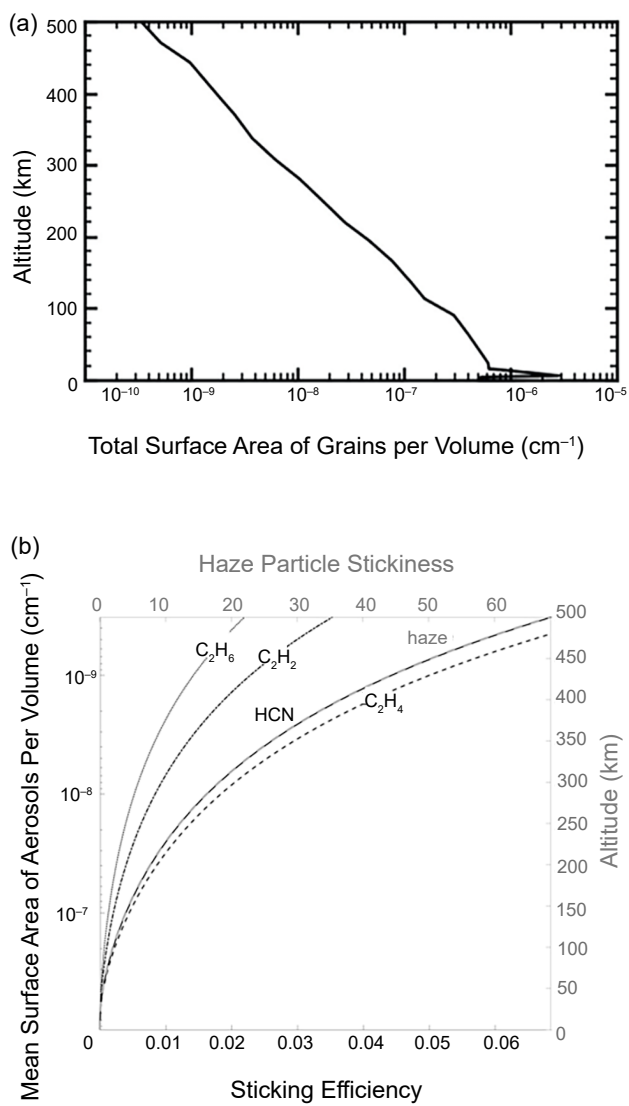

**Fig. 12. (a)** Total surface area of aerosols per volume of atmosphere from *Gao et al.* (2017). **(b)** Sticking efficiency (dimensionless) of the $C_2$ hydrocarbons and HCN as a function of haze particle surface area from *Gao et al.* (2017) and altitude (black lines), and the ability of haze particles to physically trap molecules through sticking, described as stickiness (gray dashed line). Created with data obtained from *LK2017* and *M2017*.

## 5. VARIABILITY IN PHOTOCHEMISTRY AND LONG-TERM EVOLUTION

Pluto's atmosphere could provide important clues to the origin of the planet's volatiles, but only if the evolution of Pluto's atmosphere is well understood (*Mandt et al.,* 2016; *M2017*). More specifically, its building blocks are likely to have formed in the outer solar system, but it is not clear if the temperature and pressure conditions allowed the building blocks to trap enough nitrogen in the form of $N_2$ to produce Pluto's atmosphere, or if the nitrogen was originally trapped as $NH_3$ or organics and later converted to $N_2$. Although the ratio of $N_2$ to $NH_3$ is believed to have been ~10 in the PSN (*Lewis and Prinn,* 1980), $N_2$ requires much colder temperatures to be trapped in either amorphous (*Bar-Nun et al.,* 1985, 1988) or crystalline (*Mousis et al.,* 2012, 2014) water ice. If Pluto's building blocks formed at temperatures below ~40 K, they would have accreted $N_2$ ice in greater abundance than $NH_3$ ice, but ices formed at higher temperatures would have been deficient in $N_2$. Because $NH_3$ has not been detected in the atmosphere and no upper limit has been published, we do not yet have a ratio of $N_2$ to $NH_3$ for Pluto's atmosphere.

Comets are a reasonable analog for the building blocks of Pluto because their building blocks may have formed in similar conditions. The Rosetta spacecraft provided the first detailed measurements of $N_2$ in a coma of a comet (*Rubin et al.,* 2015, 2018, 2019), providing a ratio of $N_2$ to $NH_3$ of 0.13 at Comet 67P/Churyumov-Gerasimenko (67P/C-G) (*Rubin et al.,* 2019). This ratio and the lack of detection of $N_2$ in other comets (e.g., *Cochran,* 2002) suggests that they are deficient in $N_2$ relative to $NH_3$, either as a result of their formation temperature (*Iro et al.,* 2003) or because they did not retain $N_2$ (1) beyond their first pass through the solar system (*Owen et al.,* 1993) or (2) due to internal radiogenic heating at early epochs after formation (*Mousis et al.,* 2012). The detection of $N_2$ in 67P/C-G is important because 67P/C-G is a short-period comet thought to originate in the Kuiper belt. Comparison of its composition with Pluto suggests that Pluto could have formed in a similar temperature range and may have retained some $N_2$ from the PSN. However, the greater abundance of $NH_3$ relative to $N_2$ indicates that Pluto could have obtained a larger abundance of nitrogen in its primordial ices in the form of $NH_3$.

Although organics have been proposed as a significant source (~50%) of Titan's $N_2$ (*Miller et al.,* 2019), a reasonable process for efficiently breaking the carbon-nitrogen bond in organics that is necessary for formation of $N_2$ has not been identified. Therefore, the most likely source of Pluto's nitrogen was either $N_2$ or $NH_3$ that was trapped in ices in the PSN. *Glein and Waite* (2018) showed through simple mass-balance calculations that the initial $N_2/H_2O$ ratio reported by Rosetta for 67P/C-G (*Rubin et al.,* 2015) could explain the current inventory of $N_2$ on the surface and in the atmosphere of Pluto. However, the bulk abundance of $N_2$ relative to CO in the coma of 67P/C-G is a factor of ~5 lower than the protosolar value (*Rubin et al.,* 2019), and the authors note that they were not able to explain the fact that Pluto's atmosphere and surface should have much more CO than has been detected when comparing with the $N_2$/CO ratio observed at 67P/C-G.

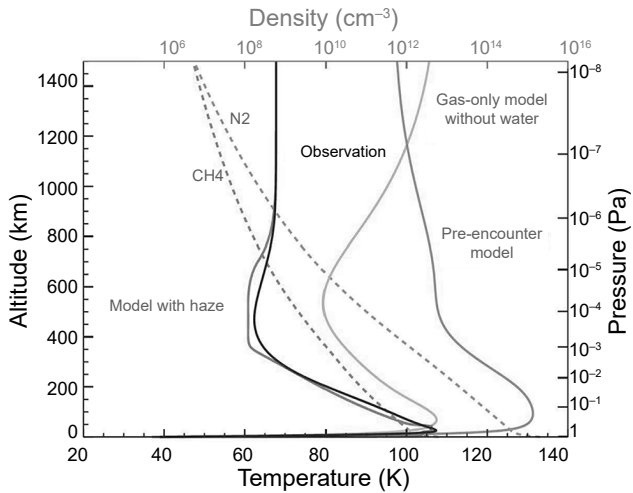

**Fig. 13.** Observed temperature and density profiles in Pluto's atmosphere (black and dashed lines), and modeled temperature profiles showing higher theoretical temperatures before the New Horizons flyby, model temperatures with atmospheric gas molecules as the only coolant, and model temperatures with haze feedback. From *Zhang et al.* (2017).

Because it is important to determine the source of Pluto's nitrogen, a valuable constraint for the origin of nitrogen would be the current nitrogen isotope ratio in $N_2$ in Pluto's atmosphere. Although this measurement cannot be taken as primordial, it can be used to constrain the origin of nitrogen if the evolution of the atmosphere is understood [e.g., *Mandt et al.* (2014) for Titan]. Once the evolution of the atmosphere is constrained, the primordial ratio derived from modeling the evolution of the atmosphere can be compared with other primordial measurements in the Sun, comets, meteorites, and giant planet atmospheres.

*Mandt et al.* (2016) and *M2017* evaluated the evolution of the nitrogen isotope ratio, $^{14}N/^{15}N$, in Pluto's atmosphere based on both pre-New Horizons information and updated based on results of the New Horizons flyby. The nitrogen isotope ratio when Pluto formed, i.e., the primordial ratio, can indicate whether the majority of Pluto's nitrogen came from $N_2$, which would have a ratio of ~450, or $NH_3$, which would have a ratio of ~130 (see *M2017* and references therein). The pre-New Horizons study found that escape of particles from the top of the atmosphere would have a dominant effect on how the isotope ratio would change over time because high escape rates in a thin atmosphere would remove more of the light isotope compared to the heavy isotope, resulting in a current ratio that is significantly lower than the primordial value. However, when applying the discovery by New Horizons that the escape rate was much lower than predicted (*G2016*) with photochemical models attempting to reproduce the ALMA upper limit for the HC$^{15}$N abundance (*L2017*), *M2017* found that photochemistry and haze feedback play a more important role in the evolution of the nitrogen isotopes over time. This is an important consideration for any studies that attempt to determine the origin of nitrogen at Pluto using future observations of the nitrogen isotopes.

## 6. SUMMARY AND CONCLUSIONS

Prior to the New Horizons flyby of Pluto, understanding of Pluto's atmosphere was limited to information gleaned from groundbased and Earth-orbiting telescopes. This information was obtained primarily by occultations of Pluto by relatively bright stars measuring the reduction in brightness of the star as Pluto passed in front of it. Based on this information, the atmosphere was known to be primarily composed of $N_2$ with trace amounts of $CH_4$ and CO, and was presumed to have hydrocarbons and nitriles based on understanding of atmospheric chemistry derived from observations of Triton's and Titan's atmospheres. Some observations suggested the possibility of haze in the lower atmosphere, but could also have been explained by a temperature gradient near the surface. Additionally, Pluto's upper atmospheric temperature was predicted to be high enough to allow for a large escape rate of $N_2$ at the top of the atmosphere.

The New Horizons flyby revolutionized our understanding of Pluto's atmosphere by showing evidence of haze layers and strong interactions between the atmospheric gas and haze particles. The observations showed that these interactions remove large amounts of HCN through condensation and $C_2$ hydrocarbons and HCN through molecules sticking to aerosols as they form and descend through the atmosphere. Physical characteristics of the aerosols can be derived from comparing models with the observations. These comparisons show that the aerosol shapes are not strictly spherical or aggregate and that their ability to adsorb molecules, or stickiness, decreases as their size increases and they age. These results are groundbreaking and will have implications for future work studying the haze in Titan's and Triton's atmospheres.

Another important outcome of the New Horizons flyby was the discovery that the temperature in the upper atmosphere was significantly lower than predicted, and that a haze feedback of cooling by radiation to space may be the cause. This has implications for the ongoing chemistry in, the escape of particles from, and the long-term evolution of Pluto's atmosphere. Reduced escape rates allow Pluto to retain volatiles on the surface and in the atmosphere long-term.

Several open questions remain to be explored by future ground- and spacebased observations as well as any future spacecraft mission to Pluto. Observations with JWST could confirm that the cooling in the atmosphere is caused by thermal radiation of aerosols. Long-term monitoring of the atmosphere using ground- and spacebased telescopes can search for evidence of haze, monitor surface pressure and methane abundance, and track temperature changes as the distance of Pluto from the Sun increases. Intentional collaboration between modeling groups is needed to resolve the differences in eddy diffusion coefficients illustrated in Fig. 7. Comparative studies of the effects of condensation and sticking to aerosols in the atmospheres of Titan, Pluto, and Triton are also needed and of high value for evaluating haze in planetary atmosphere. This will have implications for exoplanets as well. Finally, any future mission should be capable of measuring the nitrogen isotope ratio in $N_2$ to provide the needed constraints for understanding the origin on Pluto's nitrogen.